\documentclass[pre,preprint,showpacs,amsmath,amssymb]{revtex4}
\usepackage{bm}
\usepackage{graphicx}

\begin{document}

\title{Transport coefficients for dense hard-disk systems }

\author{Ram\'on Garc\'{\i}a-Rojo}
\affiliation{Institute for Computational Physics, University of
Stuttgart, Pfaffenwaldring 27, 70569 Stuttgart, Germany}

\author{Stefan Luding}
\affiliation{Technische Universiteit Delft (TUD), DelftChemTech,
Particle Technology, Julianalaan 136, 2628 BL Delft, The
Netherlands}

\author{J. Javier Brey}
\affiliation{F\'{\i}sica Te\'{o}rica,
Universidad de Sevilla, Apartado de Correos 1065, E-41080 Sevilla,
Spain}

\date{\today }

\begin{abstract}
A study of the transport coefficients of a system of elastic hard
disks, based on the use of Helfand-Einstein expressions is
reported. The self-diffusion, the viscosity, and the heat
conductivity are examined with averaging techniques especially
appropriate for the use in event-driven molecular dynamics
algorithms with periodic boundary conditions. The density and size
dependence of the results is analyzed, and comparison with the
predictions from Enskog's theory is carried out. In particular,
the behavior of the transport coefficients in the vicinity of the
fluid-solid transition is investigated and a striking power law
divergence of the viscosity in this region is obtained, while all
other examined transport coefficients show a drop in that density
range- in relation to the Enskog prediction.

\end{abstract}


\maketitle

\section{Introduction}

Transport coefficients characterize the different dissipation
mechanisms in non-equilibrium states. At the macroscopic level,
they are introduced by phenomenological equations, like the
Navier-Stokes equations for a simple fluid, which predict the time
evolution of mass, momentum and energy \cite{LyL81}. Each
transport coefficient is related to the propagation of one (or
more) of these microscopic quantities, bridging therefore the
hydrodynamic and the microscopic scale. In the case of low density
gases, the macroscopic equations have been justified, their range
of validity has been determined, and explicit expressions for the
transport coefficients have been obtained using the Boltzmann
kinetic equation as starting point \cite{Re77,McL89}. At higher
but moderate densities, the Enskog equation has also proved to
give a quite accurate description of a gas of hard spheres or
disks.

A remarkable and fundamental development in the statistical
mechanics theory of transport processes was the derivation of
expressions for the transport coefficients based on equilibrium
time-correlation functions. These are the so-called Green-Kubo
formulas, and they involve different microscopic fluxes
\cite{Zw65}. These expressions, although formal, are of general
validity and have been extensively used for the analysis and
modelling of transport in dense systems. In particular, they have
been applied to the computation of transport coefficients by means
of molecular dynamics simulations.

Alternative formal expressions for the transport coefficients are
provided by the Einstein-Helfand formulas \cite{He60,AyT93}. These
are analogous to Einstein's formula for the self-diffusion
coefficient in terms of the second moment of the displacements.
The Einstein-Helfand expressions for the other transport
coefficients involve moments of dynamical variables which are the
time integrals of the microscopic fluxes appearing in the
Green-Kubo relations.

In the last years, there has been a revived interest on transport
processes in systems composed by hard particles motivated by the
study of granular media in general, and granular gases as a
special case \cite{HHyL98,PyL01,ByP04}. The simplest model for
them is an assembly of inelastic hard spheres or disks, in which
the inelasticity is accounted for only through one constant
parameter, the coefficient of normal restitution. In the low
density limit, hydrodynamic Navier-Stokes like equations, with
explicit expressions for the transport coefficients, have been
obtained for this model, by starting from the inelastic extension
of the Boltzmann equation \cite{BDKyS98}. Moreover, it has been
shown that the transport coefficients for a dilute granular gas
can be expressed in the form of generalized Green-Kubo relations
\cite{DyB02,BDyR03}.

The Enskog equation has also been extended to inelastic particles
\cite{GyD99,Lu05}, but its density and inelasticity range of
validity is not clear. On the other hand, formally exact relations
between transport coefficients and appropriate correlation
functions, similar to the Green-Kubo and Einstein-Helfand
formulas, appear to be limited up to now to the low density limit
mentioned above and to the simplest cases of tagged particle
motion \cite{DyG01,DByL02}, although there have been some more
general proposals  \cite{GyvN00,DByB05}. Therefore, for high
densities, the only available hydrodynamic theory for granular
systems is restricted to the so-called quasi-elastic limit. The
transport coefficients are given in that limit by the same
expressions as for elastic systems, and dissipation in collisions
is taken into account only by a new term in the energy balance
equation \cite{JyS83}.

The first calculations of transport coefficients for hard-sphere
systems by means of equilibrium molecular dynamics simulations go
back to the pioneer works by Alder and coworkers
\cite{Al70a,Al70b}. The dependence of the values of the transport
coefficients on the density and also on the number of particles
used in the simulations have been analyzed. At high densities,
significant deviations from the Enskog theory predictions are
observed, especially for the self-diffusion and shear viscosity
coefficients \cite{EyW77,HyMc86,Er91,Montanero00}.

In spite of all the work done for three dimensional systems, it is
hard to find results for two dimensions, i.e., for a fluid of hard
disks. It could be argued that this is due to the presence of long
time tails in the equilibrium correlation functions appearing in
the Green-Kubo expressions of the transport coefficients, but it
must be noticed that they do not invalidate by themselves the
possibility of a hydrodynamic description \cite{Do75}. A notable
exception is ref. \cite{VyG03}, where the viscosity of a system of
hard disks is measured by using an Einstein-Helfand expression.

Since, in granular materials, a wide range of densities is
typically realized, it is useful to measure the transport
coefficients of a system of hard disks in the whole range of
densities. In this context, a global equation of state for a
system of hard disks has been proposed \cite{Lu01}. The equation
has proven to be accurate for densities ranging from the low
density region to the highest crystallization limit where caging
effects appear and free volume theories are relevant. More
precisely, it appears to be almost exact for most densities,
except for those when the system changes from a disordered state
to an ordered one. In this transition regime, memory effects
become important, hysteresis shows up, and the proposed equation
of state is only an approximate description for very slow changes
in the density.

In this paper, the transport coefficients of a system of hard
disks will be measured by means of Einstein-Helfand expressions
that are appropriate for molecular dynamics simulations with
periodic boundary conditions, using the minimum image convention
\cite{AyT93}. For continuous interaction potentials, this method
is strictly equivalent to the Green-Kubo method and has the
advantages of directly showing the positivity of the transport
coefficients and of being based on a straightforward,  numerically
robust accumulation \cite{VyG03}. In the case of hard particles,
there is a fundamental reason to employ methods based on
Helfand-Einstein expressions for the transport coefficients. The
Green-Kubo relations, except in the case of self-diffusion,
involve forces between the particles, which are ill-defined
for hard spheres or disks and there is no trivial way to extend
them to hard-particle fluids. In fact, a recent careful analysis
of the dynamics of a system of hard spheres has shown that the
correct Green-Kubo expressions for this system have a new singular
contribution due to instantaneous collisions, as well as the usual
time integral of the flux correlation functions \cite{Du02}. The
singular part vanishes in the low density limit, but gives a
relevant contribution at high densities. On the other hand, the
Einstein-Helfand formulas do not involve the forces, and have the
same form for both continuous (soft) and rigid (hard-sphere)
potentials. It must be stressed that it is an Einstein-Helfand
method that was already numerically implemented by Alder {\em et
al.} in their study of the transport coefficients in hard-sphere
fluids \cite{Al70a,Al70b}.

The most interesting finding of the present study is that the
shear viscosity shows a divergence at the crystallization density
(in a non-sheared system), while the heat conductivity is
correlated to the isotropic pressure. Thus, while pressure and
heat conductivity show a small drop at crystallization (due to the
better ordering of the particles), self-diffusion vanishes at it,
and shear viscosity diverges. This divergence is well below the
excluded volume caused divergence of pressure and heat
conductivity at the maximum density possible in hard disk systems.

The outline of the paper is as follows. In the next Section, the
Einstein-Helfand expressions for the transport coefficients are
revised, and written in a way that is appropriate for hard sphere molecular
dynamics simulations with periodic boundary conditions. The special
case of self-diffusion, in which the actual trajectories of the particles
along different cells must be followed, is first discussed. For all
the other transport coefficients, it is shown that a decomposition
of the contributions to the transport coefficients into a kinetic and
a collisional part, allows the use of the minimum image convention.
Moreover, the decomposition is especially useful for event driven simulation
algorithms. In Sec.\ \ref{s3}, the method is applied to a system of hard disks
for calculating the self-diffusion, shear viscosity, and heat conductivity coefficients
over the whole density range. The results are compared with the theoretical
predictions from Enskog's theory. Particular emphasis is put on the behavior of the
transport coefficients in the fluid-solid transition region and on characterizing
the divergence of the shear viscosity. The paper finishes with a brief discussion
of the results in Sec.\ \ref{s4}.

\section{Einstein-Helfand expressions for the transport
coefficients} \label{s2}

\subsection{The self-diffusion coefficient}
Self-diffusion is the macroscopic transport phenomenon describing
the motion of tagged particles in a fluid at equilibrium, in the
limiting case that their concentration is very low, while at the
same time they are mechanically identical to the fluid particles.
The macroscopic number density $n({\bm r},t)$ and the flux ${\bm
j}({\bm r},t)$ of tagged particles satisfy the continuity equation
\begin{equation}
\label{2.1}
\partial_{t} n({\bm r},t)+ \nabla \cdot {\bm j} ({\bm r},t)=0.
\end{equation}
The corresponding constitutive relation closing the above equation
is provided by Fick's law
\begin{equation}
\label{2.2} {\bm j}({\bm r},t)= -D \nabla n({\bm r},t),
\end{equation}
which defines the self-diffusion coefficient $D$. An expression
for this transport coefficient in terms of the second moment of
the displacements is given by the well-known Einstein formula \cite{Re77,McL89}
\begin{equation}
\label{2.3} D= \frac{1}{2d} \lim_{t \rightarrow \infty}
\frac{d}{dt} \langle |{\bm r}(t)-{\bm r}(0)|^{2} \rangle,
\end{equation}
where ${\bm r}(t)$ is the position of an arbitrary tagged particle
at time $t$, $d$ is the dimensionality of the system, and the
angular brackets denote an average over the ensemble describing the
equilibrium of the system.

To actually compute the above equation in our numerical algorithm,
two different averages have been carried out. First, an
average over the $N$ particles in the system is taken, and then a
second average over a number $\mathcal{N}$ of initial
configurations (trajectories). Assuming ergodicity of the
system, different trajectories can be generated from the same
simulation run by considering different initial times $t_{k}$.
Therefore, the full average is
\begin{equation}
\label{2.4} \langle |{\bm r}(t)-{\bm r}(0)|^{2} \rangle \frac{1}{N \cal{N}} \sum_{k=1}^{\cal{N}} \sum_{i=1}^{N} |{\bm
r}_{i}(t+t_{k}) -{\bm r}_{i}(t_{k})|^{2}.
\end{equation}
This double averaging is possible because the dynamical variable
involved in Eq.\ (\ref{2.3}) is a mono-particle property in the
present case, and all the particles are equivalent.

When using periodic boundary conditions to evaluate Eq.\
(\ref{2.4}), as in the simulations to be reported here, it is
crucial to take into account that particles originally in the
center cell in a given trajectory, have to be followed as they
cross the border of the cell. The positions in Eq. (\ref{2.4}), as writen down here, use the actual positions, but the real relative displacements should be used instead. If periodic images of the
center cell were not used, the displacements would not be obtained correctly. In this sense, the practical implementation of the
algorithm for this transport coefficient differs from those used
for the other transport coefficients to be discussed in the
following.

The simulation results for $D$ to be reported later on, will be
scaled with the value obtained from the Enskog equation in the
first Sonine approximation for $d=2$  that is \cite{Ga71}:
\begin{equation}
\label{2.5} D_{E}=\frac{1}{2 n \sigma g_{2}(\sigma)} \left(
\frac{k_{B}T}{\pi m} \right)^{1/2}.
\end{equation}
Here $T$ is the temperature, $k_{B}$ the Boltzmann constant,
$\sigma$ the diameter of the disks, $m$ their mass, and
$g_{2}(\sigma)$ the value of the equilibrium pair correlation function
at contact, which is a function of the density $n$. An estimate
for this quantity is provided by Henderson's expression
\cite{Henderson75}:
\begin{equation}
\label{2.6} g_{2}(\sigma)=\frac{1-\frac{7 \nu}{16}}{(1-\nu)^{2}},
\end{equation}
with $\nu=n \pi \sigma^{2} / 4$ being the solid fraction. The approximation in
(\ref{2.6}) is valid for densities well below the crystallization
solid fraction $\nu_c \approx 0.7$.

\subsection{Shear viscosity}
The coefficients of shear viscosity $\eta$ and bulk viscosity
$\zeta$ are defined through the macroscopic expression for the
pressure tensor ${\sf P}$ for a simple fluid \cite{LyL81}
\begin{equation}
\label{2.7} {\sf P}({\bm r},t)=p {\sf 1}-\eta \left[ \nabla {\bm
u}+ ( \nabla {\bm u})^{+} \right] + \left( \frac{2 \eta}{d} -
\zeta \right) {\sf 1} \;\; \nabla \cdot {\bm u},
\end{equation}
where $p$ is the pressure, ${\bm u}$ the flow velocity, ${\sf
1}$ the unit tensor, and the superscript $+$ means here transposed.

The Einstein-Helfand formulas for $\eta$ and $\zeta$ are
analogous to Eq.\ (\ref{2.3}). For the shear viscosity one has
\cite{He60,AyT93}
\begin{equation}
\label{2.8} \eta=\frac{m^{2}}{2Vk_{B}T} \lim_{t \rightarrow \infty
} \frac{d}{dt} \langle |G_{\eta}(t)-G_{\eta}(0)|^{2}\rangle,
\end{equation}
with $V$ the volume of the system, and
\begin{equation}
\label{2.9} G_{\eta}= \sum_{i=1}^{N} \dot{x}_{i} y_{i}.
\end{equation}

In this case, only an average over a set of different initial
conditions can be taken from the simulations, since the dynamical
variable $G_{\eta}$ already involves all the particles in the system.
Therefore, the quantity that, in principle, should be obtained from
 the simulations is
\begin{equation}
\label{2.10} \langle|G_{\eta}(t)-G_{\eta}(0)|^{2}\rangle
=\frac{1}{\mathcal{N}} \sum_{k=1}^{\mathcal{N}}  \left\{
\sum_{i=1}^{N} \left[
\dot{x}_{i}(t+t_{k})y_{i}(t_{k})-\dot{x}_{i}(t_{k})y_{i}(t_{k})
\right] \right\}^{2}.
\end{equation}
Nevertheless, this expression leads to very noisy results
and the slope of the best fit line can be only determined
with a high uncertainty. Moreover, it presents the same
difficulties as the Einstein expression for the self-diffusion
coefficient when using periodic boundary conditions, i.e., for the positions $y_i$, it requires to
follow the motion of the particles through different unit cells. It is thus
convenient to elaborate a little more Eq. (\ref{2.10}). The idea
is measuring the increments of $G_{\eta}$ for physically well defined time
intervals, instead of directly determining its actual value at successive
times.

Let us consider a time interval $\left [ t+t_{k}, t+t_{k}+\Delta t \right ]$ in
which no collisions occur in the system in a given trajectory $k$.
The purely kinetic variation of $G_{\eta}$, due to the displacements only, is
\begin{equation}
\label{2.11} \Delta G_{\eta}^{(K)} (t+t_{k})=\sum_{i=1}^{N}
\dot{x}_{i}(t+t_{k}) \dot{y}_{i}(t+t_{k}) \Delta t.
\end{equation}
In addition, there
is also a contribution due to the discontinuous change of the
velocities in collisions. Consider a collision between particles
$i$ and $j$. There is an instantaneous jump in $G_{\eta}$ given by
\begin{equation}
\label{2.12} \Delta G_{\eta}^{(C)}\dot{x}_{i}^{+}y_{i}+\dot{x}_{j}^{+}y_{j}-\dot{x}_{i}^{-}y_{i}-\dot{x}_{j}^{-
}y_{j},
\end{equation}
where we have taken into account that the positions do not change
during the instantaneous collision and the index $+$ ($-$)
indicates that the velocity is the post-collisional
(pre-collisional) one. Both velocities are related by the
collision rule for hard disks
\begin{equation} \nonumber
\dot{\bm r}_{i}^{+}= \dot{\bm r}_{i}^{-} - \dot{\bm r}_{ij} \cdot
\widehat{\bm \sigma} \widehat{\bm \sigma},
\end{equation}
\begin{equation}
\label{2.13} \dot{\bm r}_{j}^{+}= \dot{\bm r}_{j}^{-} + \dot{\bm
r}_{ij} \cdot \widehat{\bm \sigma} \widehat{\bm \sigma},
\end{equation}
with ${\bm r}_{ij}={\bm r}_{i}-{\bm r}_{j}$ and $\widehat{\bm
\sigma}$ being the unit vector joining the centers of disks $i$
and $j$ at contact and pointing from particle $j$ to particle $i$.
Using the above rule, Eq.\ (\ref{2.12}) can be rewritten as
\begin{equation}
\label{2.14} \Delta G_{\eta}^{(C)}=  y_{ij}\delta \dot{x}_{i},
\end{equation}
where $y_{ij}=y_i-y_j$, and $\delta \dot{\bm r}_{i}=- (\dot{\bm r}_{ij} \cdot
\widehat{\bm \sigma}) \widehat{\bm \sigma}$ is the change of the
velocity of particle $i$ in the collision. Since the dynamics of a
hard particle system consists of free streaming and instantaneous
collisions, Eqs. (\ref{2.11}) and (\ref{2.14}) fully determine the
time evolution of $G_{\eta}$ along a trajectory. Moreover, these
equations only involve the velocities and relative positions of
the particles. As a consequence, they avoid the difficulties of
using other Helfand-Einstein relations with periodic boundary
conditions in the simulations, as discussed by Erpenbeck
\cite{Er95}, since no contribution leads to the growth in time of the
dynamical variable $G_{\eta}$ due to the infinite
checkerboard of identical systems. This is because contributions
from pairs of particles in different unit cells cancel out
precisely due to the boundary conditions. An alternative use of a
Helfand-Einstein relation for computing the shear viscosity with
periodic boundary conditions has been discussed in \cite{VyG03}.

Equations (\ref{2.11}) and (\ref{2.14}) are particularly suitable
for event driven algorithms as the one employed in the simulations
presented in this paper. In these algorithms, the time steps are
the intervals between successive hard collisions in the system. At
every collision, the kinetic change $\Delta G_{\eta}^{(K)}$
associated with the previous time step is computed as well as the
contribution from the collision itself, $\Delta G_{\eta}^{(C)}$.
For the latter, only the positions and velocities of the pair of
colliding particles must be taken into account, while for the
kinetic contribution the motion of all the particles in the system
has to be considered.

As for the self-diffusion coefficient, the simulation results for
$\eta$ will be reported scaled with the Enskog value in the first
Sonine approximation, that for $d=2$ and densities $\nu < \nu_c$
is \cite{Ga71}
\begin{equation}
\label{2.15} \eta_{E}=\eta_{0} \left[ \frac{1}{g_{2}(\sigma)} +2
\nu +\left( 1+\frac{8}{\pi} \right) g_{2}(\sigma) \nu^{2} \right],
\end{equation}
where $\eta_{0}$ is the value in the Boltzmann limit
\begin{equation}
\label{2.16} \eta_{0}=\frac{1}{2 \sigma} \left( \frac{m k_{B}
T}{\pi} \right)^{1/2}.
\end{equation}

\subsection{Bulk viscosity}
The coefficient of bulk viscosity $\zeta$ was already introduced in Eq.\
(\ref{2.7}). The corresponding Einstein-Helfand expression is
\cite{He60,Al70a}:
\begin{equation}
\label{2.17} \zeta + \frac{4 \eta}{3}= \frac{m^2}{2 V k_{B}T}
\lim_{t \rightarrow \infty} \frac{d}{dt}
\langle|G_{\zeta}(t)-G_{\zeta}(0)- \frac{pVt}{m} |^{2}\rangle,
\end{equation}
where
\begin{equation}
\label{2.18} G_{\zeta}= \sum_{i=1}^{N} \dot{x}_{i} x_{i}.
\end{equation}
The $pVt$ term in Eq.\ (\ref{2.17}) arises from the fact that the
equilibrium average of $G_{\zeta}$ does not vanish, as it is the
case for all the other variables $G$ associated to transport
coefficients, but it is equal to the external work $pV$, defined
by the virial theorem \cite{He60,Al70a}. Since this mean value is
also computed in the simulations and it slightly shifts as the
simulations proceed, the results for the right hand side of Eq.\
(\ref{2.17}) are determined much less accurately than for the
expressions for the other transport coefficients. Additionally,
the subtraction of $\eta$, which itself is determined with some
uncertainty, causes further errors in the values estimated for
$\eta$. For these reasons, we have not been able to obtain
reliable results for the bulk viscosity in the high density
region, and no further consideration will be given to it here.
Note however, that it is still possible to split expression
(\ref{2.17}) in a kinetic and a collisional contribution, like we
have done above. For that reason, the microscopic expression of
the hydrostatic pressure is used \cite{campbell90}:
\begin{equation}
p=\frac{m N}{V} \langle\dot{x}_i \dot{x}_i\rangle + \frac{m
\sigma}{V} {\cal F} \langle \hat{x}_{ij} \delta \dot{x}_i\rangle,
\label{presscamp}
\end{equation}
where $\hat{x}_{ij}$ is the unitary vector pointing from center of particle $j$ to center of particle $i$, and $\cal{F}$ is the collision frequency \cite{Re77}. In terms of this equation and following the same reasoning as in the previous section, it is straightforward to identify that the increments on $G_{\zeta}$ in the particular case of the bulk viscosity, are related to deviations of the variable with respect to the mean value. This happens for the kinetic part,
\begin{equation}
\label{2.18b} \Delta G^{(K)}_{\zeta} (t+\Delta t) = \sum_{i=1}^{N}
\left [ \dot{x}_i(t) \dot{x}_i(t) - \langle\dot{x}_i(t)
\dot{x}_i(t)\rangle \right ] \Delta t,
\end{equation}
as well as for the collisional part:
\begin{equation}
\label{2.18c} \Delta G^{(C)}_{\zeta} (t+\Delta t) = \; \left [
x_{ij} \delta \dot{x}_i- \langle x_{ij} \delta \dot{x}_i\rangle
\right ],
\end{equation}
with $ \delta \dot{x}_i$ being the change of velocity of the
particle $i$ in its collision with particle $j$, as already
defined above. Note that expressions (\ref{2.18b}) and
(\ref{2.18c}) are again perfectly compatible with the minimum
image convection.

We finally reproduce, for the sake of completeness,  the Enskog value of the bulk
viscosity for hard disks \cite{Ga71}:
\begin{equation}
\label{2.19} \zeta_{E}= \frac{8 \nu^{2} g_{2}(\sigma)}{\pi \sigma}
\left( \frac{m k_{B} T}{\pi} \right)^{1/2}.
\end{equation}

\subsection{Thermal conductivity}
The coefficient of thermal conductivity $\lambda$ is defined by
the Fourier law for the heat flux ${\bm q}({\bm r},t)$
\cite{LyL81}
\begin{equation}
\label{2.20} {\bm q}({\bm r},t)= - \lambda  \nabla T({\bm r},t),
\end{equation}
and the Einstein-Helfand expression for it is \cite{HyMc86}
\begin{equation}
\label{2.21} \lambda=\frac{1}{2Vk_{B}T^{2}} \lim_{t \rightarrow
\infty} \frac{d}{dt} \langle
|G_{\lambda}(t)-G_{\lambda}(0)|^{2}\rangle,
\end{equation}
with
\begin{equation}
\label{2.22} G_{\lambda}=\sum_{i=1}^{N} x_{i} e_{i}.
\end{equation}
Here $e_{i}$ is the energy of particle $i$,
\begin{equation}
\label{2.23} e_{i}=\frac{m \dot{\bm r}_{i}^{2}}{2}.
\end{equation}
As it was the case with Eq.\ (\ref{2.10}), also Eq.\ (\ref{2.21}) is
not appropriate for numerical simulations with periodic boundary
conditions, since it involves the positions of the particles.
Therefore, we are going to transform it in a similar way, i.e.,
measuring separately the increments of the dynamical variable
$G_{\lambda}$ between collisions and the collisional
contributions. The variation of $G_{\lambda}$ along a trajectory
$k$ in a time interval $\Delta t$ such that no collisions take
place is
\begin{equation}
\label{2.24} \Delta G_{\lambda}^{(K)}= \frac{1}{2} \sum_{i=1}^{N}
\left[
\dot{x}_{i}(t+t_{k})e_{i}(t+t_{k})-\dot{y}_{i}(t+t_{k})e_{i}(t+t_{k})
\right],
\end{equation}
where the isotropy of the equilibrium state has been taken into
account. In this way, the statistical quality of the simulation
results is increased.

In a collision between particles $i$ and $j$, the instantaneous
change in $G_{\lambda}$ is
\begin{eqnarray}
\label{2.25} \Delta G_{\lambda}^{(C)}& = &
x_{i}e_{i}^{+}+x_{j}e_{j}^{+}-x_{i}e_{i}^{-}-x_{j}e_{j}^{-}
\nonumber \\
& = & x_{ij}(e_{i}^{+}-e_{i}^{-}) =  x_{ij} \delta e_i.
\end{eqnarray}
The second equality follows from the total energy conservation in
the collisions here considered. Using again the isotropy of the system, we can write
\begin{equation}
\label{2.26} \Delta G_{\lambda}^{(C)}= \frac{
(x_{ij}+y_{ij})\delta e_i}{2}\, .
\end{equation}
Equations (\ref{2.24}) and (\ref{2.26}) do not contain absolute
positions of the particles, but only their relative positions and
their velocities. Therefore, we have obtained again an expression
that does not require to follow the track of the particles all
along the simulation, namely to use the so-called {\em unfolded}
particle coordinates. The values of the position and velocities of
the particles expressed in {\em folded} coordinates can be used,
without taking care of the cell crossing.

Enskog's  theory prediction for the heat conductivity is
\begin{equation}
\label{2.26a}
\lambda_E= \lambda_{0} \left[ \frac{1}{g_{2}(\sigma)} + 3 \nu + \left ( \frac{9}{4} +
\frac{4}{\pi} \right ) \nu^{2} g_{2}(\sigma) \right],
\end{equation}
where $\lambda_{0}$ is the Boltzmann  value
\begin{equation}
\label{2.26b}
\lambda_{0}=\frac{2 k_{B}}{\sigma} \left( \frac{k_{B}T}{\pi m} \right)^{1/2}.
\end{equation}

\section{Results}
\label{s3} In this Section, results from several series of event
driven simulations are presented. We consider an homogeneous,
freely evolving system of elastic granular disks with periodic
boundary conditions, implemented by means of the minimum image
convention \cite{Lu92}. Systems with different numbers of
particles and volume have been simulated, and the dependence of
the transport properties on density and particle number has been
investigated. A typical simulation started with a square lattice
of particles having a Gaussian velocity distribution. After a
transient period, the system reaches a equilibrium homogeneous
state. Then, the measurement of the different properties of
interest was carried out, using the procedure described in the
previous section. For every system, the above process was repeated
a number of times, typically $300$, in order to generate the
ensemble average over different trajectories. Averages over different initial times were also considered for overriding
the lack of statistical precision in the cases of the shear
viscosity and the thermal conductivity, as compared with the
self-diffusion coefficient. That implied performing
longer simulations and store and handle the numerical data. The
number of initial times used was in all cases larger than 200.

\subsection{Self-diffusion coefficient}
\begin{figure}
\begin{center}
\includegraphics[scale=0.5,angle=-90]{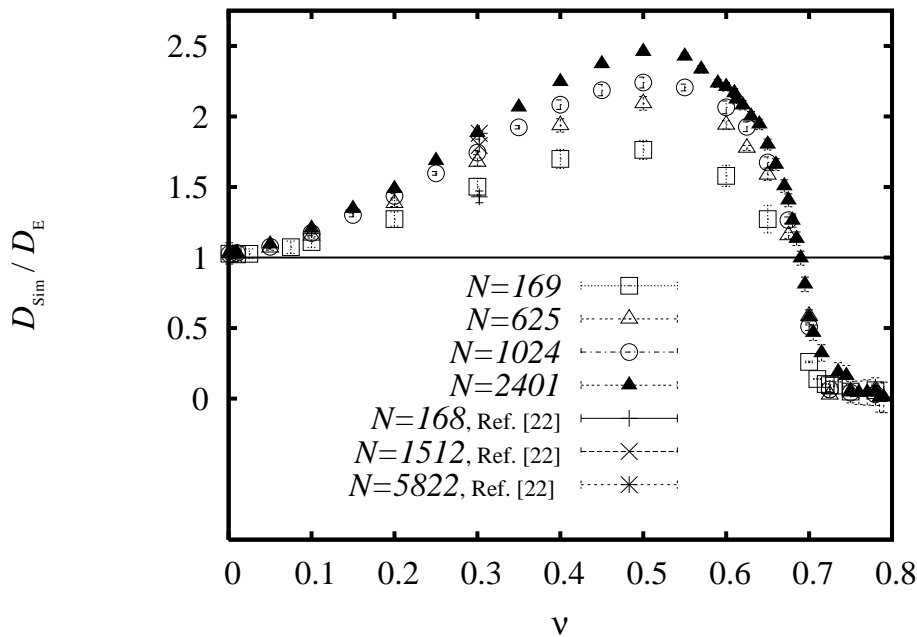}
\end{center}
\caption{Scaled self-diffusion coefficient for a hard-disk fluid,
as a function of the packing fraction $\nu$. For each value of the
density, systems of $N=2401$, $N=1024$, $N=625$, and $N=169$
identical particles have been considered. Also included are
previous results obtained by Holian {\em et al.} \cite{EyW77},
using non-equilibrium simulation techniques. Note that the
$D_{sim}=D_E$ for $\nu=0.69$. The extrapolated tangent of the
curve at this point crosses the line $D_{sim}=0$ at $\nu \approx
0.705$, a value which is fairly close to $\nu_c$. } \label{fig1}
\end{figure}

In Fig. \ref{fig1}, the results obtained for the self-diffusion
coefficient, $D_{sim}$, as a function of the solid fraction $\nu$
are presented. For each density, systems with different numbers of
particles $N$ have been considered, namely $N=169,625,1024$, and
$2401$. Moreover, the reported values are the average over 300
independent trajectories. A strong deviation of the Enskog value
$D_{E}$ is observed, even for relatively low densities. Also shown
are some previous results obtained by Holian {\em et al.}
\cite{EyW77} at $\nu=0.3$ using a non-equilibrium molecular
dynamics method. They consistently agree with the results being
reported here. For large enough packing fractions $\nu$, the
particles become trapped in a crystalline lattice and no free
movement is possible \cite{AyW62}. Therefore, the self-diffusion
coefficient must vanish in this limit. This explains the rather
fast decay to zero observed in the simulations. Of course, these
high density effects are not captured by the Enskog equation.

The series of values of $D$ obtained for each value of $N$ in the
interval $ 0\leq \nu \leq 0.5$ have been fitted to a third degree
polynomial
\begin{equation}
\label{3.1} \frac{D_{sim}(\nu)}{D_{E}(\nu)}= a+b \nu + c \nu^{2} +d
\nu^{3}.
\end{equation}
The values of the fitting coefficients are given in Table
\ref{table1}. There, it is seen that the coefficient $a$,
characterizing the dilute limit, seems to be weakly dependent on
the number of particles used in the simulation, and clearly larger
than unity, indicating that the dilute limit is slightly
underestimated by the expression for $D_E$ we have used. Moreover,
we have carried out simulations with different boundary conditions
and found always the same value consistently. In fact, similar
deviations have been previously observed \cite{BRCyG00}. This can
be due to the fact that the Enskog expression given by Eq.\
(\ref{2.5}) has been computed in the first Sonine approximation,
as already mentioned. It is possible that the consideration of
higher order polynomial corrections in the Sonine expansion would
improve the agreement between theory and simulations in the low
density limit.
\begin{table}[htb]
\begin{center}
\begin{tabular}{|c||c|c|c|c|}
\hline
N&a&b&c&d\\
\hline \hline
$169$& $1.0207$ & $0.0433$ & $8.629$ & $-11.43$ \\
\hline
$625$& $1.0299$ & $0.3367$ & $9.775$ & $-12.39$ \\
\hline
$1024$& $1.0259$ & $0.4683$ & $10.711$& $-13.48$\\
\hline
$2401$& $1.0287$ & $0.5971$ & $11.930$ & $-14.75$\\
\hline
\end{tabular}
\end{center}
\caption{Empirical fit of the simulation results for the
self-diffusion coefficient to the third order polynomial in Eq.\
(\ref{3.1}). The error of the fitting for each value is of the order of the last figure given.}\label{table1}
\end{table}

\begin{figure}
\begin{center}
\includegraphics[scale=0.5,angle=-90]{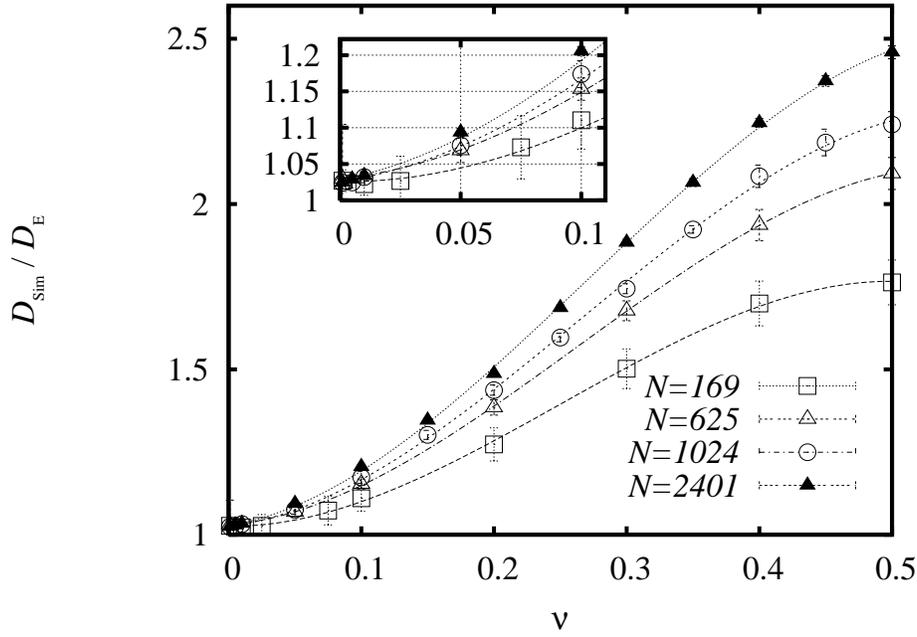}
\end{center}
\caption{Dependence of $D_{sim}$ on the solid fraction $\nu$ for
low and moderate densities. The symbols are some of the
simulation results given in Fig.\ \ref{fig1}. The lines are
the fits to the polynomial (\ref{3.1}) with the coefficients
given in Table \ref{table1}. In the inset,
the low density region is enlarged.}
\label{fig2}
\end{figure}

\begin{figure}
\begin{center}
\includegraphics[scale=0.5,angle=-90]{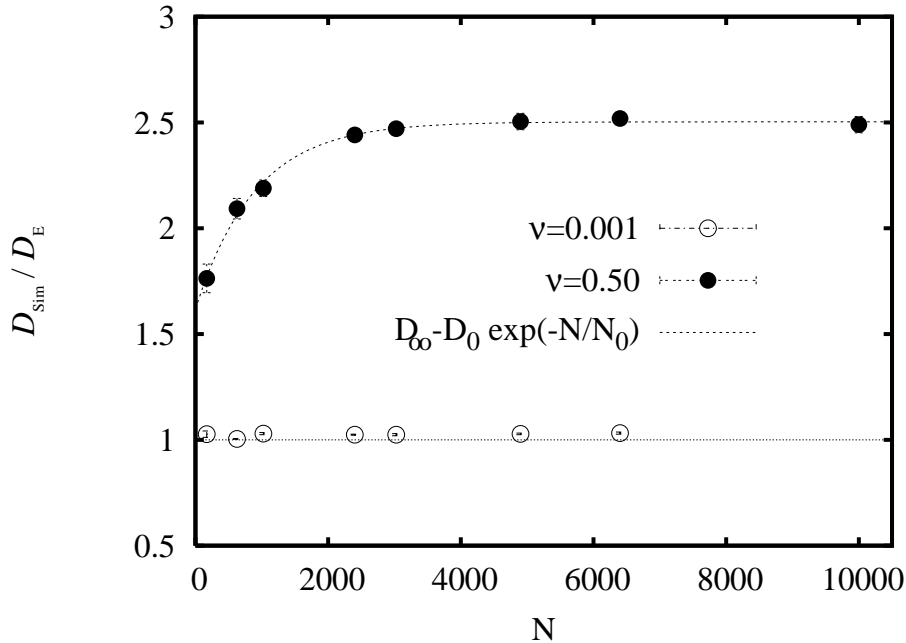}
\end{center}
\caption{Dependence of the measured self-diffusion coefficient on
the number of particles $N$ for systems with densities
$\nu=0.001$ and $\nu=0.5$, respectively. The results have been averaged over
30 trajectories for the system with $N=10000$ particles, and over 300 trajectories for
all the other systems. The error bars in this scale are smaller than the symbols used.}
\label{fig3}
\end{figure}

Figure \ref{fig1} and Table \ref{table1} clearly indicate a strong
dependence of the measured value of the self-diffusion coefficient
on the system size. Of course, it is expected that the simulation
results converge to a well defined value as the number of
particles increases, although this is not at all clear from Fig.
\ref{fig1}, especially for densities around $\nu = 0.5$. To check
this convergence and characterize it, two series of simulations
corresponding to $\nu=0.001$ and $\nu=0.5$, respectively, have
been performed. The results, as functions of the number of
particles used, are presented in Fig. \ref{fig3}. It is seen that
in the low density case, accurate size-independent results are
already obtained with a small number of particles, namely with
$N=169$. On the other hand, for $\nu=0.5$ the convergence is much
slower, and reliable results require a few thousands of particles.
More precisely, the dependence on $N$ in this cases is quite well
fitted by an exponential function $D_{\infty}- D_{0}\exp
(-N/N_{0})$, with $N_{0}=900$. In any case, the existence of an
asymptotic value of $D_{\infty}$ follows clearly from the above
results.

\subsection{Shear viscosity}
\begin{figure}
\begin{center}
\includegraphics[scale=0.5,angle=-90]{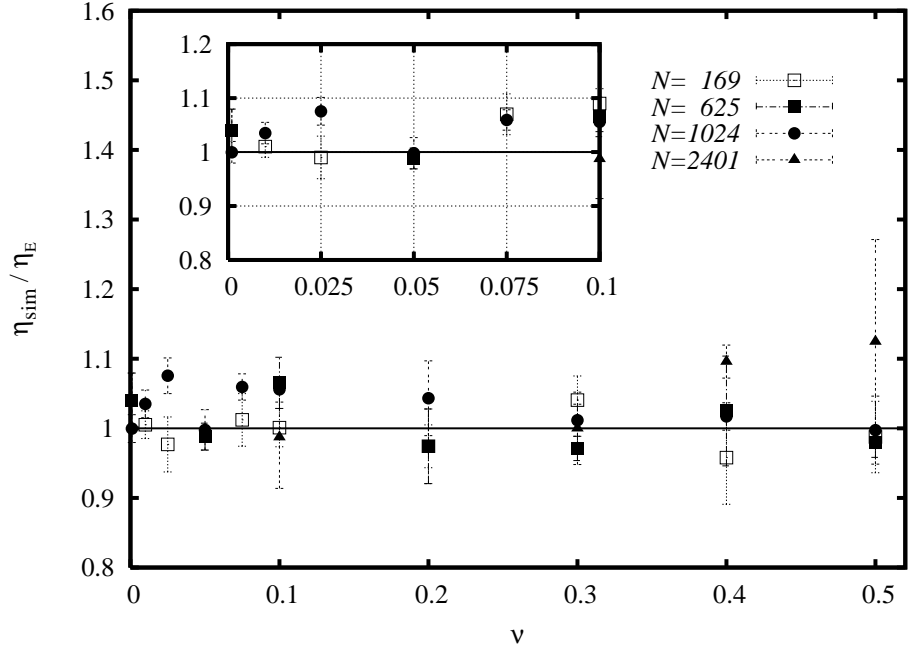}
\end{center}
\caption{Shear viscosity coefficient normalized with the Enskog prediction
as a function of solid fraction for dilute and  moderately  dense systems. The symbols
are from the simulations with different number of particles $N$. Averages
over 300 different initial conditions and over 300 different initial times
have been taken. The low density region is amplified in the insert.}
\label{fig4}
\end{figure}

The simulation results for the shear viscosity at low and
intermediate densities are shown in Fig. \ref{fig4}. The
deviations from Enskog theoretical predictions is typically under
$10\%$  of the value for low densities, and this extends up to the
transition to the ordered state. This is true even for the
simulations with the lowest number of particles. In fact, no
strong dependence of the measured value of the transport
coefficient on $N$ can be inferred from the simulation data in
this range. A similar behavior was found by Alder {\em et al.} for
a system of hard spheres \cite{Al70a}.

Enskog theory clearly underpredicts shear viscosity in the range $
0.5 < \nu <0.68$, contrary to what was found in the case of the
self-diffusion coefficient, which dropped at $\nu_c$. This is so
because the collision frequency in that range of densities is
overestimated by Enskog's theory. In the crystalline region, the
measurement of the shear viscosity becomes rather difficult, since
the expected linear behavior of the increment in time of the
dynamical variable $G_{\eta}(t)$ disappears (see Eq.\
(\ref{2.8})).

\begin{figure}
\begin{center}
\includegraphics[scale=0.5,angle=-90]{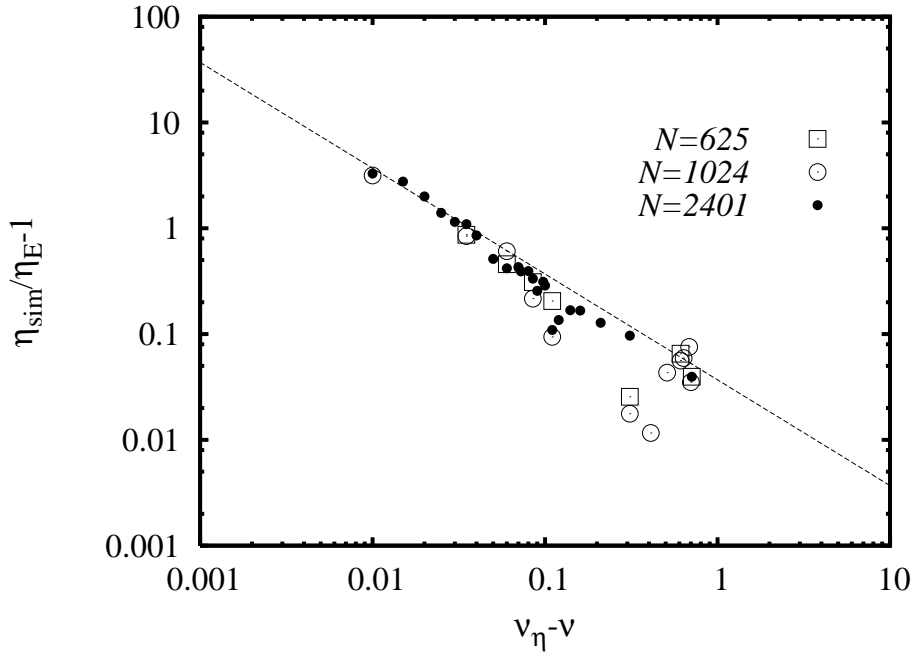}
\end{center}
\caption{Divergent behavior of the shear viscosity coefficient.
The symbols show simulations with different number of particles,
as indicated. They have been obtained by averaging as in Fig\ \ref{fig4}.
The dashed line is the power-law fitting given in Eq.\ (\ref{3.2}).}
\label{fig5}
\end{figure}

In Fig. \ref{fig5} the deviations of the measured values of
$\eta_{sim}$ from Enskog theory in the high density region are
plotted in a logarithmic scale.  A power-law divergent behavior of
the viscosity is observed as the density approaches the critical
value. Moreover, no shift of critical (viscosity) density
$\nu_{\eta}$ is observed as the number of particles increases. The
dashed line in the figure is the function
\begin{equation}
\label{3.2} \eta^* (\nu) = c (\nu_{\eta}-\nu)^{-1},
\end{equation}
with $c=0.037 \pm 0.001$, and $\nu_{\eta}=0.71 \pm 0.01$. This latter
value approximately agrees with the density for which the
self-diffusion coefficient vanishes (see Fig.\ \ref{fig1}), and
also with the critical (crystallization) density in the global
equation of state proposed in \cite{Lu01}. Let us mention that for
$N=169$ no linear behavior of $G_{\eta}$ was found for densities
$\nu \agt 0.65$ and, therefore, no results with this number of
particles are included. The behavior of $\eta_{sim}$ in this
region, in fact, seems to depend on $N$, as can be observed in the
figure.

\begin{figure}
\begin{center}
\includegraphics[scale=0.5,angle=-90]{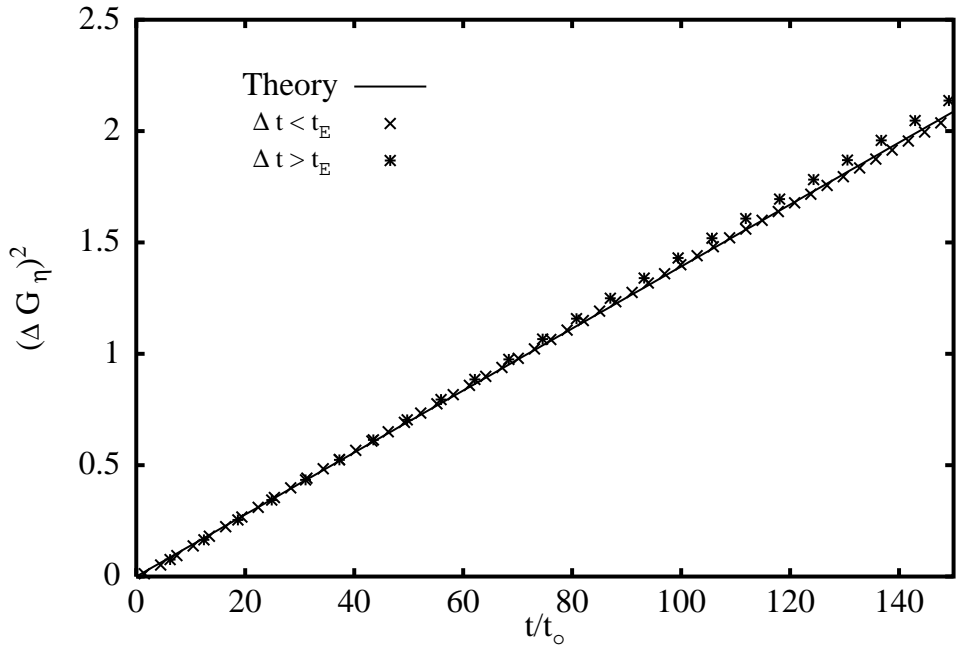}
\end{center}
\caption{Time evolution of the dynamical variable $( \Delta
G_{\eta})^2$ obtained with two different time intervals $\Delta
t$, one shorter than the mean collision time $t_{E}$ and the other
larger than it. The number of particles is $N=1024$ and the
density $\nu=0.30$, in both cases. The solid line is the Enskog
prediction. Time is scaled by $t_0$, Boltzman's mean time between
collisions. Note that $t_0 = t_E$ in the dilute limit, where
$g_2(\sigma) \rightarrow 1$. } \label{fig6}
\end{figure}

It is worth stressing here the relevance of making the
measurements of the corresponding dynamical variable in the
simulations with a frequency higher than the collision frequency
in the fluid. In this case, the validity of the numerical
procedures discussed in the previous section is guaranteed and the
results obtained from them can be expected to be correct. On the
other hand, if the time interval between successive measurements
is increased with respect to the mean collision time of the
system, the results for the time evolution of the corresponding
dynamical variable $G$, may not present a linear behavior. But,
even if it turns out to be linear, the slope can lead to wrong
values of the corresponding transport coefficient. As an example,
a comparison of the measurement of $G_{\eta}$ in two simulations
made with the same system of $N=1024$ particles and density
$\nu=0.30$ is presented in Figure \ref{fig6}. More accurate
results are obtained with a time step between measurements of
$G_{\eta}$ shorter than the mean time between collisions. This is
a consequence of the effect of multiple collisions occurring
between successive measurements of $G_{\eta}$, which invalidates
the arguments leading to Eqs.\ (\ref{2.11}) and (\ref{2.14}).
Although this applies, in principle, to both the shear viscosity
and the thermal conductivity, the simulation results show that the
influence of the time step between the measurements employed is
stronger for the former than for the latter.

\subsection{Thermal conductivity}
\begin{figure}
\begin{center}
\includegraphics[scale=0.5,angle=-90]{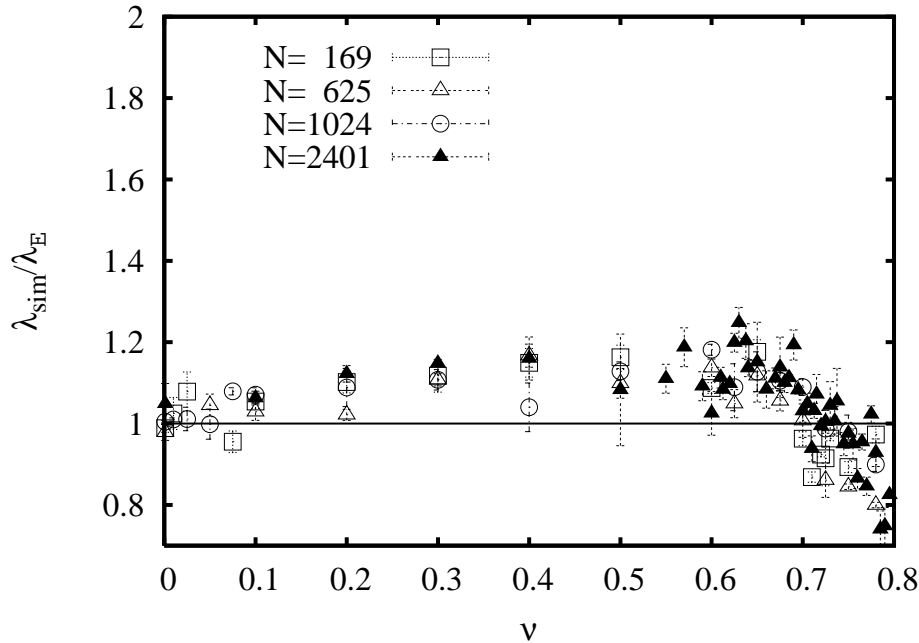}
\end{center}
\caption{Thermal conductivity coefficient as a function of the solid fraction
$\nu$ for an elastic hard-disk systems of different numbers of particles $N$, as indicated.
The values have been averaged in the same way as in Fig. \ref{fig4}. Error-bars are
shown when larger than the symbols used.}
\label{fig7}
\end{figure}

Fig.\ \ref{fig7} depicts the results obtained for the heat conductivity. Although the discrepancies with the Enskog
predictions are not large (they are of the order of 10\%), the
qualitative behavior as a function of the density resembles that
of the self-diffusion coefficient reported in Fig.\ \ref{fig1}. It
exhibits a maximum around $\nu \simeq 0.55$, decaying below the
Enskog prediction for larger densities ($\nu \agt 0.7$). This
decay is due to the decreased mean free path due to the ordering
of the particles. Of course, in contrast with $D_{sim}$,
$\lambda_{sim}$ does not vanish in the ordered region, since there
is still considerable transport of energy through collisions.
Moreover, $G_{\lambda}$ was found to exhibit linear behavior in
the transition to the ordered state, even for the smallest system
considered ($N=169$). Similarly to the case of the shear
viscosity, no systematic dependence of the results on the number
of particles used is observed. Let us mention that, although
small, the deviations from the Enskog predictions in Fig.
\ref{fig7} are larger than those found by Alder {\em et al.} for a
system of hard spheres \cite{Al70a}.

\begin{figure}
\begin{center}
\includegraphics[scale=0.5,angle=-90]{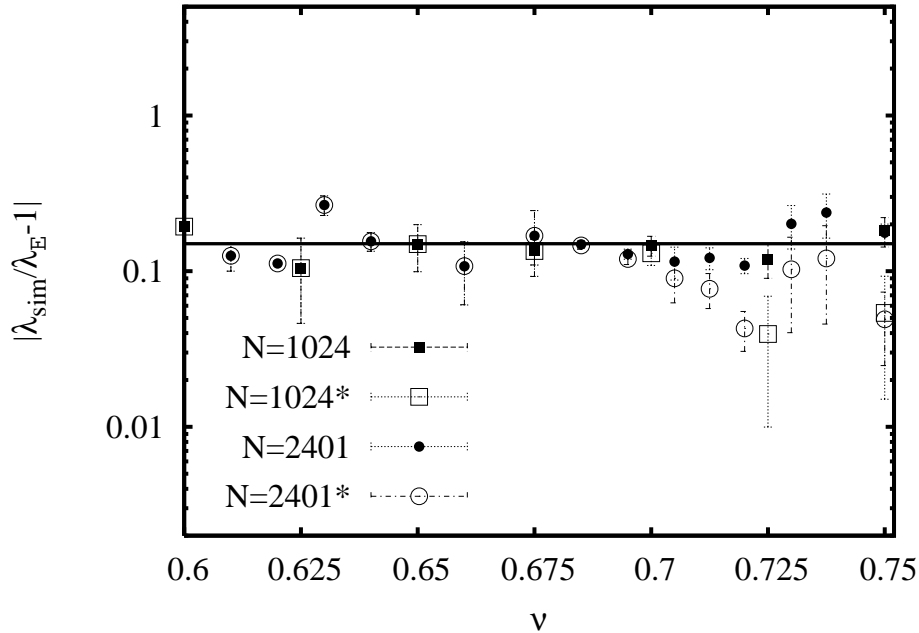}
\end{center}
\caption{Deviation of the values measured for the thermal
conductivity coefficient, $\lambda_{sim}$, from Enskog's value,
$\lambda_E$, as a function of the solid fraction $\nu$ for dense
systems. The values have been averaged in the same way as in Fig.
\ref{fig4}. For the scaling of the tagged series $N=1024^{*}$ and
$N=2401^{*}$, the values $\nu_{c}=0.71$ and $m_0=0.01$ have been
used. More details about the scaling of these results are given in
the text. The solid line indicates the fixed value
$(\lambda_{sim}/\lambda_E -1)=0.15$.} \label{fig8}
\end{figure}

In Fig.\ \ref{fig8} we investigate the transition region
($\nu>0.65$), where the dispersion of our measurements is clearly
higher. The deviation of the measurements with respect to Enskog's
value is plotted using two different expressions for the pair
correlation function $g_2(\sigma)$ in equation (\ref{2.26a}). On
the one hand, we have used the formula given in equation
(\ref{2.6}), and these are the open symbols. For the other two
series (solid symbols), a semi-empirical formula as given in
reference \cite{Lu01}, valid in the whole range of densities
studied here, was used:
\begin{equation}
g_{2}^{L}(\sigma)=g_2(\sigma)+m(\nu|\nu_{c},m_0) \left [
\frac{P_{dense}(\nu)}{2 \nu}-g_2(\sigma) \right ],
\label{(34)}
\end{equation}
where $m(\nu | \nu_{c},m_0)$ is a connecting function, and
$P_{dense}$ is the reduced pressure in the dense region (see Ref.
\cite{Lu01} for more details). The explicit expressions of these
functions are next given in terms of the parameters $\nu_c$,
$\nu_{\eta}$ and $m_0 \approx 0.0111$:
\begin{equation}
m(\nu | \nu_{c},m_0)= \frac{1}{1+\exp \left[ -(\nu-\nu_{c})/m_0
\right]}\, ,
\label{(35)}
\end{equation}
\begin{equation}
P_{dense}(\nu) \approx \frac{2 \nu_{\eta}}{\nu_{\eta}-\nu} \left [ 1 -0.04
(\nu_{\eta}-\nu)+3.25(\nu_{\eta}-\nu)^3 \right ].
\end{equation}

In Fig. \ref{fig8} results for high values of $\nu$ are plotted.
The log-log representation is used in order to be consistent with
Figure \ref{fig5}. The empirical pair correlation function
$g_{2}^{L}$ is expected to work better for values of the density
above $\nu \approx \nu_{c}=0.70$ \cite{Lu01,Lu02,Lu04}.  The
deviation from the theoretical value remains approximately
constant for a wide range of solid fractions, covering the low and
moderate densities. When $g_{2}^{L}$ is used, the deviation
remains constant even beyond the transition to the dense
configuration, while there is a clear deviation of the data if the
simplified form of equation (\ref{2.6}) is used. This is more
clearly observed in the range $(0.65 \alt \nu \alt 0.74)$. The
effect of the correction for smaller values of $\nu$ is
nevertheless negligible.

It has been suggested \cite{HyMc86} that the good quantitative agreement between the
simulation results and the Enskog prediction for the heat conductivity, may be attributed
to the absence of a significant ``tail'' in the autocorrelation function appearing in its
Green-Kubo expression. Nevertheless, the Helfand-Einstein approach of the
transport coefficients we have used, is free from the practical difficulties associated
with those tails. Moreover, the fact that a linear time dependence of the relevant Helfand moment has been found for both the shear viscosity and the heat conductivity, seems
to confirm the validity of the formulas used.

\section{Discussion}
\label{s4}

In this paper, the transport coefficients of a system of elastic
hard disks have been evaluated by means of equilibrium molecular
dynamics simulations. In order to avoid the fundamental
difficulties recently identified in the use of the standard
Green-Kubo formulas in hard-particle systems \cite{Du02}, a
Helfand-Einstein representation has been employed. The
consideration of periodic boundary conditions in the simulation of
a non-sheared, isotropic, homogeneous, freely evolving system of
elastic hard spheres or disks forces some modifications of the
usual Einstein-Helfand's formulas for the transport coefficients,
especially if the minimum image convention is used. Moreover, the
expressions proposed here are especially suitable for event driven
methods. They allow a detailed  study of the dependence of the
coefficients on the system size and density.

For the self-diffusion coefficient, $D$, the Enskog approximation
leads to values that underestimate the simulation results by
factors up to two, for moderate values of the density ($\nu \leq
0.3$), the discrepancies being already relevant at rather low
densities. The observed density dependence of the transport
coefficient is well fitted by a third order polynomial for $\nu
\leq 0.5$, with coefficients that slightly depend on the number of
particles, $N$, of the system. It has been verified that $D$ tends
to a well defined limit as $N$ becomes large enough. At higher
densities, the transition liquid-solid is clearly depicted in the
behavior of the self-diffusion coefficient. It rapidly falls to
zero as a consequence of the caging of the particles. Finite size
effects are more relevant for dense systems, in which the
self-diffusion coefficient approaches its asymptotic value
exponentially with $N$.

For the shear viscosity the dependence of the results on the size
of the system is much smaller. Also much weaker deviations from
the Enskog prediction are observed at low and intermediate
densities. Nevertheless, closer to the gas-solid transition, a
power law divergent behavior has been identified. Interestingly,
the density value for which the viscosity would become eventually
singular ($\nu_c \simeq 0.71$), agrees with the density at
which the system begins to show an ordered triangular structure
\cite{TyCh88,LyS00,TTyD00}.

Note that the results here presented
refer to ``non-sheared'' systems. We have avoided therefore the
problem of the system becoming inhomogeneous and developing a
shearband \cite{alam05}. A sheared system will not show the
divergence found for the viscosity because of the shearbanding
instability.

\begin{figure}
\begin{center}
\includegraphics[scale=0.5,angle=-90]{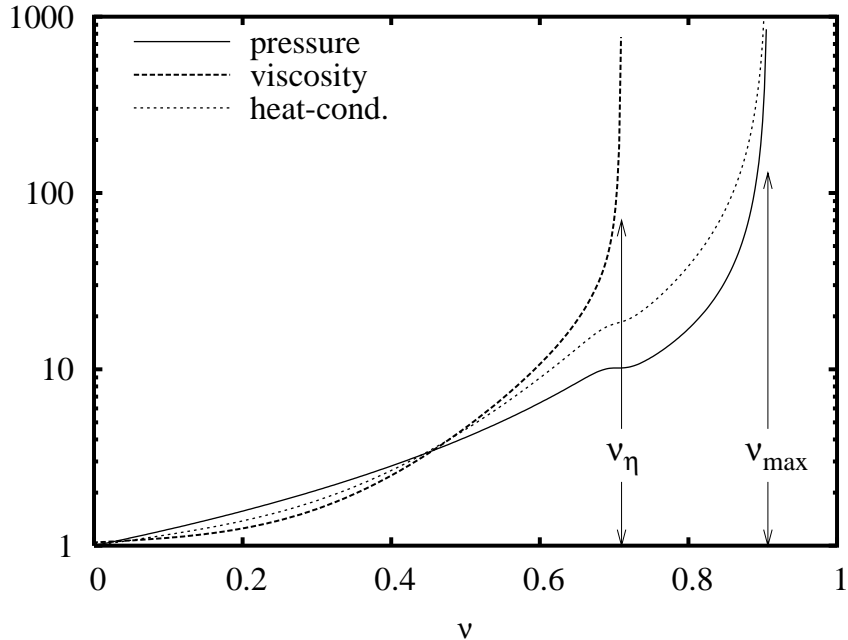}
\end{center}
\caption{Schematic plot of the transport coefficients.
The solid line gives the pressure $P=1+2 \nu g_L$, with $g_L$ from 
Eq. (\ref{(35)}); the dashed line gives the scaled shear viscosity, i.e.\ 
Eq.\ (\ref{2.16}), where $g_L$ is used instead of $g_2$, and multiplied
by our empirical correction factor $1+c_\eta$ in Eq.\ (\ref{(34)}) ; and the dimensionless heat condictivity from Eq.(31), also with $g_L$ used instead of $g_2$.}
 \label{fig9}
\end{figure}

The pressure also diverges but at a considerably higher 
density $\nu_{max} \approx 0.9069$.  At the crystallization density
$\nu_c \approx \nu_\eta \approx 0.7$, both pressure and heat
conductivity show a drop relative to the Enskog prediction due to the better ordering in the crystalline phase (see Fig. \ref{fig9}). The use of a more elaborated expression of the pair correlation function, valid in a wider range of densities than Henderson's approximation, improves the agreement of our data with Enskog theory. There were no data obtained for the shear viscosity above $\nu_{\eta}$.

In conclusion, we have found Enskog theory working  rather well
for pressure, heat-conductivity and shear viscosity well below the 
crystallization density $\nu_c$.  When the Enskog expressions are
corrected by an appropriate pair correlation function $g_L$, which
accounts for the better ordering in the crystalline phase, the theory
performs well for pressure and heat-conductivity up to the maximal 
possible density $\nu_{max}$.  

Only the shear viscosity shows a power-law divergence at $\nu_\eta \approx \nu_c$ with values above Enskog theory already becoming visible at intermediate densities. Thus, shear viscosity behaves differently than the other transport coefficients studied. Its divergence, implying that the shear modes are hindered for $\nu > \nu_\eta$. This could in fact be understood as one reason for shear-band formation. A sheared system at high densities typically splits into  shear bands (with lower density) and a compressed dense crystal (with correspondingly higher density). From a different point of view, our observations are also consistent with the concept of dilatancy: A dense packing with $\nu > \nu_\eta$ can only be sheared by first experiencing dilatancy so that $\nu$ drops below $\nu_c$.

\section{Acknowledgments}
The research of J.J.B. was supported by the Ministerio de
Educaci\'on y Ciencia (Spain) through Grant No.\ FIS2005-01398
(partially financed by FEDER funds). S.L. acknowledges helpful
discussion with J.T.Jenkins and M. Alam, as well as financial
support of the DFG (Deutsche Forschungsgemeinschaft, Germany) and
FOM (Stichting Fundamenteel Onderzoek del Materie, The
Netherlands) as financially supported by NWO (Nederlandse
Organisatie voor Wetenschappelijk Onderzoek).

\end{document}